\documentclass[12pt]{article}
\usepackage{graphicx}
\usepackage{cite}

\let\oldbibliography\thebibliography
\renewcommand{\thebibliography}[1]{%
  \oldbibliography{#1}%
  \setlength{\itemsep}{0pt}%
}

\makeatletter
\renewcommand\section{\@startsection {section}{1}{\z@}%
                                   {-3.5ex \@plus -1ex \@minus -.2ex}%
                                   {2.3ex \@plus.2ex}%
                                   {\normalfont\large\bfseries}}
\renewcommand\subsection{\@startsection{subsection}{2}{\z@}%
                                     {-3.25ex\@plus -1ex \@minus -.2ex}%
                                     {1.5ex \@plus .2ex}%
                                     {\normalfont\normalsize\bfseries}}
\renewcommand\subsubsection{\@startsection{subsubsection}{3}{\z@}%
                                     {-3.25ex\@plus -1ex \@minus -.2ex}%
                                     {1.5ex \@plus .2ex}%
                                     {\normalfont\normalsize\bfseries}}
\makeatother

\newcommand\pubnumber{}
\newcommand\pubdate{}

\def\institute{Physik Department T31, Technische Universit\"at
  M\"unchen, James-Franck-Str. 1, D-85748 Garching, Germany}
\def\support{\footnote{Work supported by the Alexander von Humboldt
    Foundation, in the framework of the Sofja Kovalevskaja Award
    Project ``Event Simulation for the Large Hadron Collider at High
    Precision''.  }}

\def\Title#1{\begin{center} {\Large #1 } \end{center}}
\def\Author#1{\begin{center}{ \sc #1} \end{center}}
\def\Address#1{\begin{center}{ \it #1} \end{center}}

\newcommand\pubblock{\rightline{\begin{tabular}{l} \pubnumber\\
         \pubdate  \end{tabular}}}
\newenvironment{Abstract}{\begin{quotation}  }{\end{quotation}}
\newenvironment{Presented}{\begin{quotation} \begin{center} 
             PRESENTED AT\end{center}\bigskip 
      \begin{center}\begin{large}}{\end{large}\end{center} \end{quotation}}

\def\beq{\begin{equation}}
\def\eeq#1{\label{#1}\end{equation}}
\def\eeqn{\end{equation}}

\def\beqa{\begin{eqnarray}}
\def\eeqa#1{\label{#1}\end{eqnarray}}
\def\eeqan{\end{eqnarray}}

\let\bar=\overbar

\def\Dslash{\not{\hbox{\kern-4pt $D$}}}
\def\dslash{\not{\hbox{\kern-2pt $\del$}}}

\def\msb{{\bar{\ssstyle M \kern -1pt S}}}

\begin{document}
\begin{titlepage}
\pubblock

\vfill
\Title{Accurate predictions for t-channel single top-quark production}
\vfill
\Author{Rikkert Frederix\support}
\Address{\institute}
\vfill
\begin{Abstract}
Single top-quark production through the exchange of a t-channel
W-boson is the main electroweak production mode for top quarks at the
Large Hadron Collider. I discuss a recent development in the
theoretical predictions for this process, in which t-channel
single-top with an additional jet is combined, at NLO accuracy, with
t-channel single top production, without the introduction of a merging
scale.
\end{Abstract}
\vfill
\begin{Presented}
$11^\mathrm{th}$ International Workshop on Top Quark Physics\\
Bad Neuenahr, Germany, September 16--21, 2018
\end{Presented}
\vfill
\end{titlepage}
\def\thefootnote{\fnsymbol{footnote}}
\setcounter{footnote}{0}

\section{Introduction}
At the Large Hadron Collider (LHC) running at 13 TeV, the second
largest production mode for top quarks is t-channel single-top quark
production, being close to one third of the pair production
mode~\cite{Aliev:2010zk,Kant:2014oha,Kidonakis:2010ux,Kidonakis:2013zqa}. At
variance with the pair production mode, which is governed by the
strong interaction, t-channel single-top production is mediated by a
W-boson and is therefore an electroweak (EW) process. This make it
sensitive to a different class of new physics
scenarios~\cite{Tait:2000sh,Cao:2007ea,Atwood:2000tu,Drueke:2014pla,Aguilar-Saavedra:2017nik,Zhang:2016omx}
and directly sensitive to the CKM matrix
element $|V_{tb}|$~\cite{Alwall:2006bx,Lacker:2012ek,Cao:2015doa,Alvarez:2017ybk}.

Within the Standard Model, this process is known at the next-to-leading
order (NLO) in the strong coupling for a long
time~\cite{Harris:2002md,Campbell:2004ch,Cao:2005pq,Campbell:2009ss,Campbell:2009gj,Campbell:2012uf}
and recently also the next-to-NLO corrections have been
computed~\cite{Brucherseifer:2014ama,Berger:2016oht,Berger:2017zof}. Beyond
fixed-order perturbation theory, effects from all order resummation
have been studied as
well~\cite{Wang:2010ue,Kidonakis:2011wy,Cao:2018ntd}. For the
Monte-Carlo simulation of completely exclusive predictions, the NLO
corrections have been matched to parton
showers~\cite{Frixione:2005vw,Frixione:2008yi,Frederix:2012dh,Alioli:2009je,Bothmann:2017jfv,Papanastasiou:2013dta,Jezo:2015aia,Frederix:2016rdc}
and these NLOPS predictions are what is currently being used by the
experimental collaborations to simulate the single-top
process. Finally, also the NLO corrections in the EW have been
considered in
refs.~\cite{Beccaria:2008av,Bardin:2010mz,Frederix:2018nkq}.

While the NLOPS predictions have a remarkable accuracy, for single-top
production they do not include the latest developments regarding
multi-jet
merging~\cite{Alioli:2011nr,Hamilton:2012np,Hoeche:2012yf,Frederix:2012ps,Platzer:2012bs,Alioli:2012fc,Lonnblad:2012ix,Hamilton:2012rf,Alioli:2013hqa,Frederix:2015fyz,Bellm:2017ktr}. The
results presented here, which are based on
ref.~\cite{Carrazza:2018mix}, show that this limitation has now been
lifted. 

\section{Method}
The method used to combine NLO single-top together with NLO single-top
plus jet production is the MINLO procedure~\cite{Hamilton:2012np}
together with (improved and refined) ideas presented in
ref.~\cite{Frederix:2015fyz} to recover the NLO accuracy of the lowest
multiplicity single-top process.

The MINLO method starts from a fixed-order NLO computation for
single-top plus jet production, $d\sigma_{{\scriptscriptstyle
    \mathrm{NLO}}}^{{\scriptscriptstyle \mathrm{STJ}}}$. This process
was implemented in the POWHEG BOX
framework~\cite{Alioli:2010qp,Alioli:2010xd} as the STJ
generator~\cite{Carrazza:2018mix} using matrix elements generated with
Madgraph4 and
MadGraph5\_aMC@NLO~\cite{Campbell:2012am,Alwall:2014hca}. The
single-top plus jet calculation is not reliable for observables that
are inclusive over the extra jet; indeed, a strict fixed-order
prediction will result to an infinitely large cross section when the
energy or transverse momentum of the extra radiation approaches
zero. Indeed, in this limit the cross section will be dominated by
logarithms of the differential jet rate, $\sqrt{y_{{\scriptscriptstyle
      12}}}$, over the hard scale of the process.  To cure this
infinity a simple all-order resummation is added (at leading
logarithmic accuracy). Schematically this results in the MINLO cross
section being defined as
\begin{equation}
  d\sigma_{{\scriptscriptstyle \mathcal{M}}}=%
  \Delta(y_{{\scriptscriptstyle 12}})\,%
  \left[\,%
    d\sigma_{{\scriptscriptstyle \mathrm{NLO}}}^{{\scriptscriptstyle \mathrm{STJ}}}%
    -%
    \left.%
    \Delta(y_{{\scriptscriptstyle 12}})%
    \right|_{\bar{\alpha}_{{\scriptscriptstyle \mathrm{S}}}}%
    \,d\sigma_{{\scriptscriptstyle \mathrm{LO}}}^{{\scriptscriptstyle \mathrm{STJ}}}\,%
    \right]\,,%
\end{equation}
where $\Delta(y_{{\scriptscriptstyle 12}})$ is the Sudakov form
factor. The dampening of the Sudakov form factor cures the infinity,
making the cross section finite even when looking inclusively over all
radiation.  The Sudakov form factor can be computed order-by-order in
perturbation theory, in which the lowest order coefficients are
universal and depend only on the flavor of the particles involved in
the core process, while at higher orders more process dependent
information enters. The term proportional to the
$\mathcal{O}(\alpha_S)$ expansion of the form factor,
$\left.\Delta(y_{{\scriptscriptstyle
    12}})\right|_{\bar{\alpha}_{{\scriptscriptstyle \mathrm{S}}}}$, is
included to make the NLO expansion of the MINLO cross section
identical to the original NLO single-top plus jet cross section. It
can be shown~\cite{Hamilton:2012np} that by including only the universal
coefficients LO accuracy for inclusive observables can be obtained:
\begin{equation}
  \int dy_{{\scriptscriptstyle 12}} \frac{d\sigma_{{\scriptscriptstyle
        \mathcal{M}}}}{y_{{\scriptscriptstyle
        12}}}=d\sigma_{{\scriptscriptstyle
      \mathrm{LO}}}^{{\scriptscriptstyle \mathrm{ST}}}
  +\mathcal{O}(\alpha_S).
  \label{only-lo}
\end{equation}
To also get the NLO terms correct, more information needs to be added
to the Sudakov form factor. This can either be done by using analytic
expressions~\cite{Hamilton:2012rf,Luisoni:2013kna,Hamilton:2016bfu} or
by numerical fits~\cite{Frederix:2015fyz,Carrazza:2018mix} for the
higher order corrections that enter the Sudakov form factor. For the
t-channel single-top process it is the latter approach that has been
pursued.

\subsubsection*{Fitting procedure}
In order to improve the agreement in eq.~(\ref{only-lo}) to NLO
accuracy, we state that the following must hold
\begin{equation}
  \frac{d\sigma_{{\scriptscriptstyle
        \mathrm{NLO}}}^{{\scriptscriptstyle
        \mathrm{ST}}}}{d\Phi}\,=\,\int
  dy_{12}\,\frac{d\sigma_{{\scriptscriptstyle \mathcal{M}}}}{d\Phi
    dy_{12}}\,\delta\Delta(y_{12})\,,
  \label{nlo}
\end{equation}
where $\delta\Delta(y_{12})$ is the modification needed to the original Sudakov form factor and has the form
\begin{equation}
  \ln\delta\Delta(y_{12})\,=\,-2\int_{y_{12}}^{Q_{bt}^{2}}\frac{dq^{2}}{q^{2}}\,\bar{\alpha}_{{\scriptscriptstyle
      \mathrm{S}}}^{2}\,\mathcal{A}_{2}(\Phi)\,\ln\frac{Q_{bt}^{2}}{q^{2}}\,.
  \label{delta-form}
\end{equation}
The $\mathcal{A}_{2}(\Phi)$ is an unknown function of $\mathcal{O}(1)$
that we need to obtain by imposing the equality in eq.~(\ref{nlo})
numerically. $\Phi$ is the complete Born phase-space and is therefore
four-dimensional. However, one dimension is trivial (rotation about
the collision axis), hence there remains a three-dimensional fit to be
done. At the most basic level, in order to achieve our accuracy goal,
the only requirement that need be satisfied by the modifications
indicated in eqs.~(\ref{nlo}) and (\ref{delta-form}), is that
$\mathcal{A}_{2}(\Phi)$ be $\mathcal{O}(1)$; this is demonstrated
formally, and numerically, in ref.~\cite{Carrazza:2018mix}.
Since we do not want to introduce a functional bias and the
function to fit is inside an exponent, we train an artificial neural
network to form $\mathcal{A}_{2}(\Phi)$ that minimizes the following loss
function
\begin{equation}
  \mathcal{L} \,\, = \,\, \sum_{i=1}^{N_{\rm bins}} \, %
  \left[\, %
  \sum_{j=1}^{N} \, w^{\scriptscriptstyle {\rm ST}}_{i,j} %
  \, - \,%
  \sum_{k=1}^{N'} \, %
                   w^{\scriptscriptstyle{\rm STJ}}_{i,k} \,%
                   \mathrm{e}^{\,%
                     \tilde{\mathcal{A}_{2}}(\Phi_{i})%
                     \,%
                     \mathcal{G}_{2}(\lambda) %
                              }%
  \,\right]^2
\end{equation}
after having discretized the phase-space in $N_{\rm bins}$ bins. $N$
($N'$) is the number of ST (STJ) events used in carrying out the fit.
$w^{\scriptscriptstyle{\rm ST(J)}}_{i,j}$ is the weight of the $j$th
ST(J) event in bin $i$ of the discretized Born variable parameter
space. $\tilde{\mathcal{A}_{2}}(\Phi)$ is the neural network
prediction for the desired effective Sudakov form factor coefficient
(and $\mathcal{G}_{2}(\lambda)$ a known function). A meta-search was
carried to find the optimal network architecture and using 25 million
ST events and 18 million STJ events the 42 independent parameters of
the resulting network where fitted. It was found that the model
prediction for $\tilde{\mathcal{A}_{2}}(\Phi)$ is of $\mathcal{O}(1)$
(or smaller) all over the phase-space (as it must be for the procedure
to be reliable), apart from some extreme corners where somewhat larger
values where found, possibly due to a lack of statistics in the
fitting procedure. The obtained model has been made available as part
of the STJ generator in the POWHEG BOX.

\section{Results}
In the following we discuss a limited set of results that validate the
new STJ$^\star$ predictions. For this, we compare it to the existing
ST generator~\cite{Alioli:2009je}, as well as the new STJ
generator~\cite{Carrazza:2018mix}. The latter includes the MINLO
procedure, but without the non-universal, numerically fitted
coefficients in the Sudakov form factor that make the STJ$^\star$
prediction NLO accurate for ST observables, i.e., with
$\delta\Delta(y_{12})=0$. The events have been generated for the
13~TeV LHC, using NNLO NNPDF3.0 parton
distributions~\cite{Ball:2014uwa} and a top mass of 172.5~GeV. The
events have been showered with Pythia8~\cite{Sjostrand:2014zea},
including the alternative momentum reshuffling for initial-final QCD
dipoles~\cite{Cabouat:2017rzi}. The top quarks are kept stable
throughout the simulation and no hadronization effects or other
non-perturbative modeling have been included.

The layout of the plots is the following. In the main panel the
absolute cross sections are given, with the estimates from the ST
simulation in green, the STJ in blue and the best predictions,
STJ$^\star$ in red. The three insets display ratios of the various
results to one another. The bands display the renormalization and
factorization scale dependence, generated by varying the two scales
independently by a factor two up and down, discarding the
contributions in which the scales are varied in opposite
directions. The small darker red band variation in STJ$^\star$
corresponds to a factor 4 variation (up and down) of the hard scale,
$Q_{bt}$, that enters the $\delta\Delta(y_{12})$ contribution, see
eq.~(\ref{delta-form}).

\begin{figure}[htb]
\centering
\includegraphics[height=8cm]{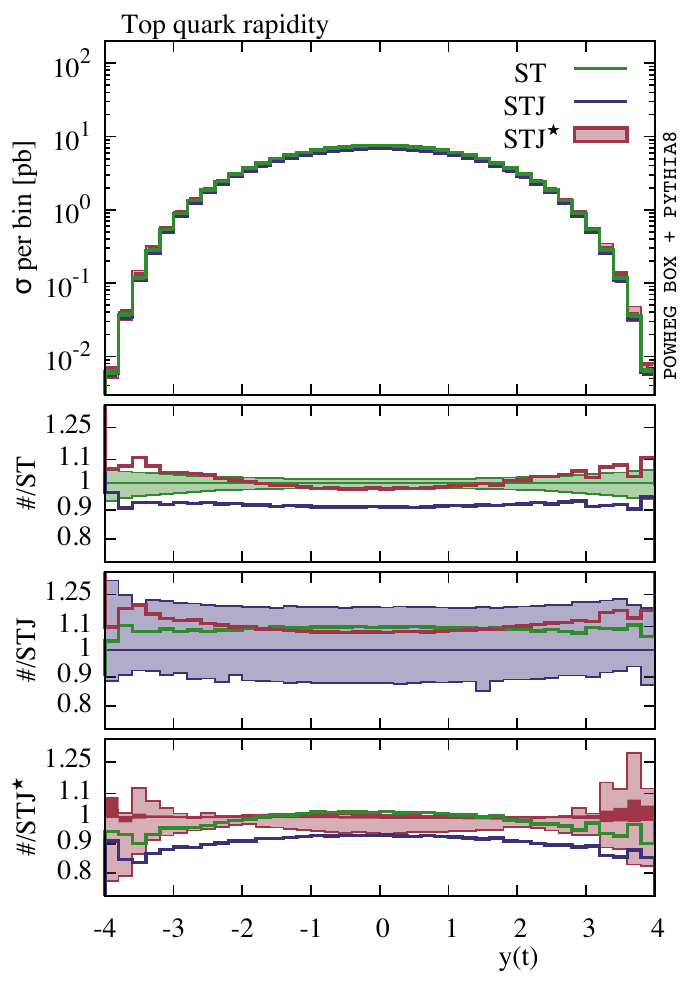}
\includegraphics[height=8cm]{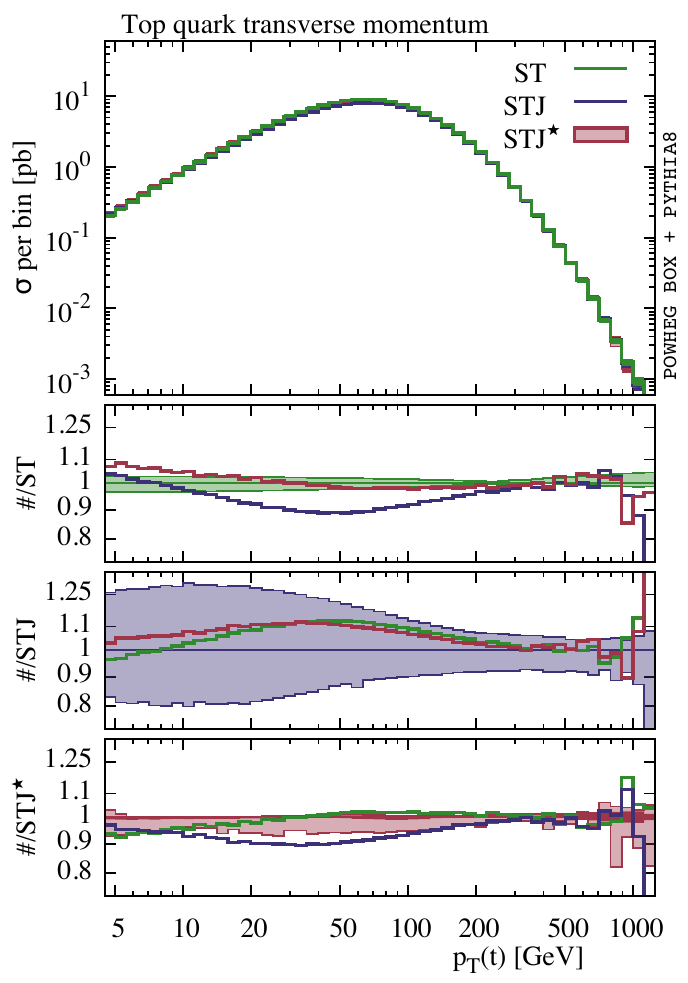}
\caption{Predictions for the rapidity (left) and transverse momentum
  (right) of the top quark.}
\label{fig1}
\end{figure}

In fig.~\ref{fig1} the rapidity (left) and the transverse momentum
(right) of the top quark are plotted. These observables are described
at NLO accuracy with the ST generator, since they are inclusive over
all extra radiation. For these observables the STJ calculation is only
LO accurate. As expected the STJ$^\star$ predictions agree within the
uncertainty band with the ST predictions. This shows that including
the $\delta\Delta(y_{12})$ into the STJ generator, i.e., updating it
STJ$^\star$, indeed recovers NLO accuracy for inclusive observables.

\begin{figure}[htb]
\centering
\includegraphics[height=8cm]{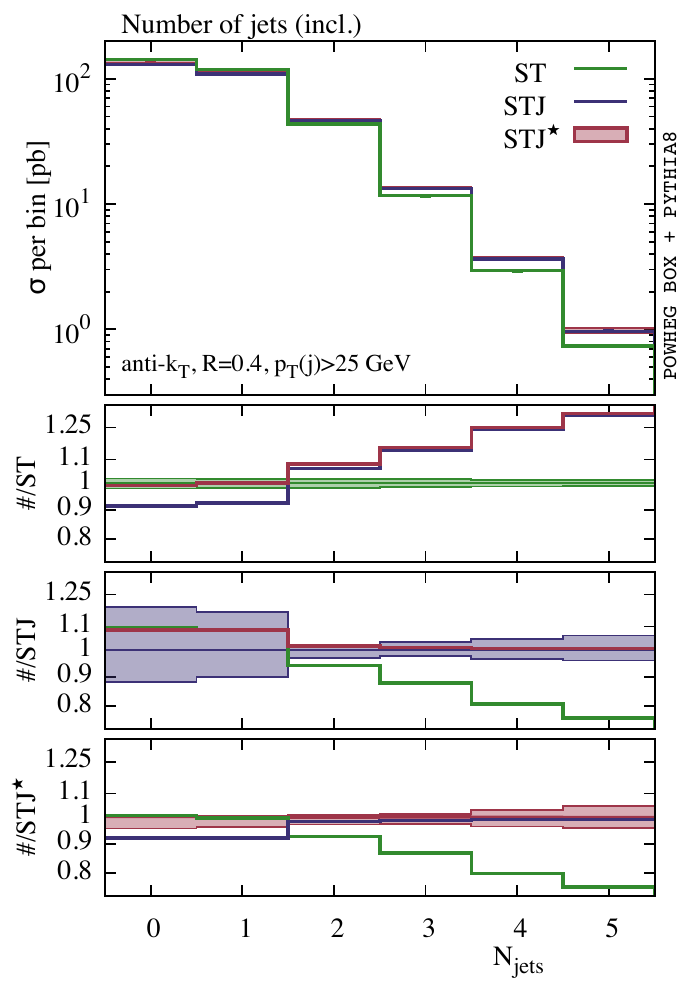}
\includegraphics[height=8cm]{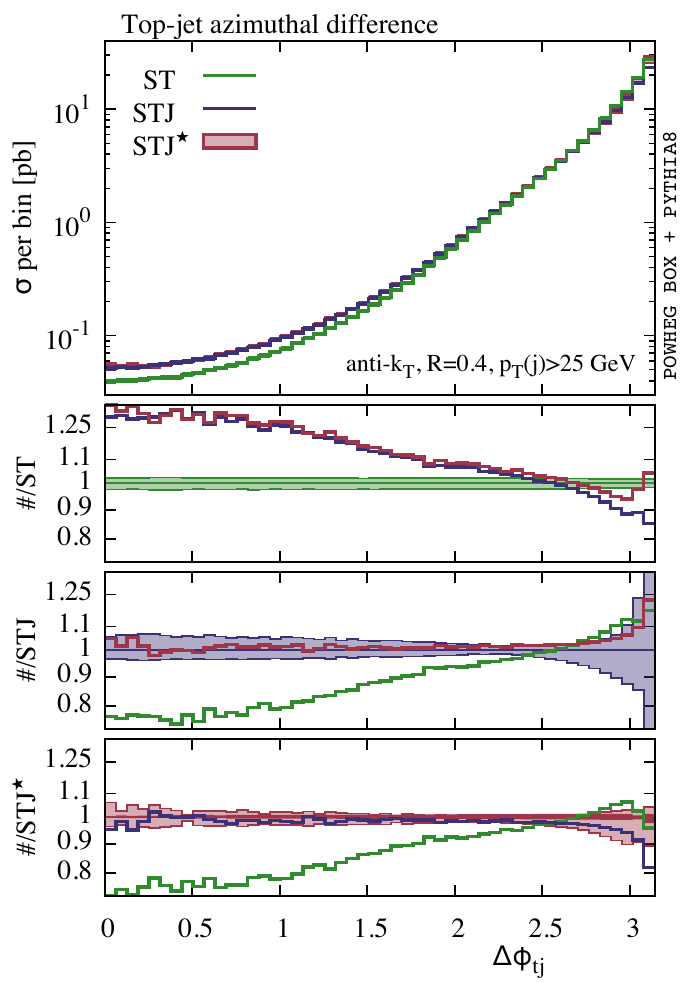}
\caption{Predictions for the jet multiplicity (left) and azimuthal
  angle between the top quark and the hardest jet (right).}
\label{fig2}
\end{figure}

In fig.~\ref{fig2} the jet multiplicity (jets are defined with the
anti-$k_{t}$ algorithm with the radius parameter set to $R=0.4$) and
the azimuthal angle between the top quark and the hardest jet are
shown in the left and right plots, respectively. In the left plot, it
is directly obvious that where the ST generator is NLO accurate (in
the 0 and 1-jet bins) the STJ$^\star$ generator agrees with that
predictions, while where the STJ generator is preferred, it agrees
with the latter. Note that the rather small scale dependence band for
the higher multiplicity bins is a well-understood artifact of the
POWHEG method, that underlies all these predictions. For small values
of the top quark-jet azimuthal separation (right hand plot), there
must be at least one additional hard object in the event, hence this
observables needs the STJ matrix elements to be NLO accurate. Indeed,
the STJ$^\star$ agrees with STJ in this region of phase-space. On the
other hand, where the top quark and the jet are in opposite
directions, the STJ and STJ$^\star$ predictions start to differ, with
the latter moving closer to the ST predictions. Indeed, in this
region of phase space there cannot be a (single) hard object in
association with the top quark and the hardest jet, hence, ST is
preferred over STJ as the most accurate prediction.

\section{Conclusions}
The new STJ$^\star$ predictions give a consistent description of ST
and STJ observables at NLO accuracy. It is based on the MINLO method,
but with a numerical fit for additional coefficients in the Sudakov
form factor to make the STJ generator also NLO accurate for
observables inclusive over the extra jet. This makes the STJ$^\star$
agree STJ where STJ is NLO accurate and with ST where the latter is
NLO accurate. The new STJ process, together with the numerical model
to enhance it to STJ$^\star$, have been made available in the POWHEG
BOX framework.

\subsubsection*{Acknowledgements}
I would like to thank Stefano Carrazza, Keith Hamilton and Giulia
Zanderighi for the collaboration on the work presented here.

\begin{footnotesize}

\end{footnotesize}

\end{document}